# Meta-Metrics for Simulations in Software Engineering on the Example of Integral Safety Systems

Christian Berger[3], Delf Block[2], Christian Hons[2],
Stefan Kühnel[1,2], André Leschke[2], Bernhard Rumpe[1], Torsten Strutz[2]

[1]Software Engineering
RWTH Aachen University, Germany
rumpe@se-rwth.de

[2]Volkswagen Aktiengesellschaft
Entwicklung Fahrzeugsicherheit
stefan.kuehnel@volkswagen.de

[3]Department of Computer Science and Engineering
Chalmers | University of Gothenburg, Sweden
christian.berger@gu.se

**Abstract:** Vehicle's passengers and other traffic participants are protected more and more by integral safety systems. They continuously perceive the vehicle's environment to prevent dangerous situations by e.g. emergency braking systems. Furthermore, increasingly intelligent vehicle functions are still of major interest in research and development to reduce the risk of accidents. However, the development and testing of these functions should not rely only on validations on proving grounds and on long-term test-runs in real traffic; instead, they should be extended by virtual testing approaches to model potentially dangerous situations or to re-run specific traffic situations easily. This article outlines *meta-metrics* as one of today's challenges for the software engineering of these cyber-physical systems to provide guidance during the system development: For example, unstable results of simulation test-runs over the vehicle function's revision history are elaborated as an indicating metric where to focus on with real or further virtual test-runs; furthermore, varying acting time points for the same virtual traffic situation are indicating problems with the reliability to interpret the specific situation. In this article, several of such meta-metrics are discussed and assigned both to different phases during the series development and to different levels of detailedness of virtual testing approaches.

## 1 Introduction and Motivation

Today's vehicles are more and more equipped with sensor- and actuator-based driver assistance systems which may be subdivided into so called comfort- and safety-oriented systems [NCA12, BR12b]. These functions realized by such systems and integrated in automobiles as well as the higher demand on computational power increase the complexity level of software and cross-linkage. The modular concept of AUTOSAR has been established to manage the development of integral safety systems (ISS) and to standardize interfaces between different abstraction layers within a device controller. This enables potentials of specialization for both OEM and suppliers regarding soft- and hardware development and the optimization of this process itself. However software with a high level of complexity requires new methods in testing because firstly, a failure or malfunction within safety-critical functions might lead to disastrous consequences. Secondly, the actual situation of the environment detected by sensors and further interpreted by an algorithm may be subject of much higher variation than air-conditioning or audio systems. But the development process should also focus on usability in respect of the customer's requirements, and thus, the quality assurance for a function like predictive pedestrian protection should clearly rely on objective metrics but should also consider the appealing of the function's behavior.

Thus, there has to be an adequate consideration of subjective aspects during the development process which is done by, of course, intensive test-drives throughout the test and application phase. In this case "application" means finding an ideal set of parameters to get a comprehensive appealing of a function with focus on behavior and impression. Nevertheless it is a difficult task to make those subjective aspects more objective and to transfer these steps into a virtual environment assuring the function's quality [BR12a]. Another possibility to enhance quality of software is to identify certain failures even earlier than their occurrence in former projects. This can be done by designing and applying meta-metrics, which allow an analysis and in-depth evaluation of existing metrics that measure a function or specific software module. These meta-metrics will operate independently from a specific system and test context like a concrete simulation environment and may be means for a more general purpose to tackle one of today's software engineering challenges for cyber-physical systems [Ber12].

The remainder of this article is organized as follows: Sec. 2 summarizes existing publications to this topic and related work. Sec. 3 defines the term "meta-metrics" focusing on the automotive domain. In Sec. 4, some meta-metrics are applied exemplarily on an abstracted ISS test process. Sec. 5 discusses experiences and results and gives an outlook on future work.

## 2 Related Work

The topic of using meta-metrics to support the development of safety functions especially in the automotive domain has not been investigated fully to the best knowledge of the authors. However, the field of metrics in computer science and software engineering exists for a longer period, so there are many standard references, which concentrate on discussing particular software metrics or how to derive metrics within a given context. Such references are e.g. [LDA97, FP98, EFR07, Lig09, SSB10], which consider software quality related processes to assure quality and software measurement in general. An introduction into this topic is given by [Glo05] and [Aly06].

Recent research aims in particular for identifying and systemizing new metrics on the one hand and tries to evaluate as many of those to provide guidance to software engineers in choosing the right ones for a project on the other hand. A survey of object-oriented metrics present Xenos et al. in [XSZC00] who gathered traditional software metrics applicable to object-oriented contexts as well as those specifically designed for object-oriented environ-ments. This collection has been evaluated by meta-metrics, which were partly existent to that time, and which partly had to be evolved for this survey. Meta-metrics have been used to enable the assessment of a specific selection of basic metrics in a better way.

An analogical approach is presented by [SX09]. There, certain metrics get characterized by various categories, which simultaneously form the meta-metrics to improve the selection of measures for the evaluation of e-commerce systems. In this case Stefani also aims for supporting stakeholders by giving them an orientation, *where*, *which one*, and *for what purpose* metrics are supposed to be used. Some examples for these meta-metrics are "measurement scale", "measurement independence", or "accuracy".

Woodings addresses in [Woo99] the need for measuring evaluation results within the scope of the development process in software projects to identify potential improvements and proposed two meta-metrics by converting existing metrics: The first one is DeMarco's Estimating Quality Factor, which focuses on the rapid convergence to an accurate figure during a project, and the second one is a definition to provide a lower boundary on errors for multiple initial estimates. He shows that both are able to achieve the requirements for metrics and their usefulness.

Baroni and Abreu present in [BBeA03] a formal library for aiding metrics extraction (FLAME) with the purpose of formalizing object-oriented design metrics definitions. To that end, they use the Object Constraint Language (OCL) as part of the Unified Modelling Language (UML) [OMG12] upon its meta-model and combine several thereof to "functions" included into their library. They evaluated various design models to verify the usefulness of that methodology. A catalogue of formalized metric definitions is also proposed.

In [WN10], the authors reveal the importance of meta-measurement approaches to derive challenges in measurement, e.g. minor significance of a single metric in contrast to a selection, subjectivity of measurement, and the different views of a measured system. Thus, they define certain requirements for those approaches as stability and understandability, which need to be fulfilled by a measurement, or the dynamic expendability for measures. Weber and Nimmich present an evaluation meta model and thereof its derivable meta-measures that they validate in the context of services such as web-based serices and other.

The authors of [BRR$^+$10b, BRR10a] showed how metrics are able estimate the potential to form product lines from existing legacy software. Therefore, they firstly describe a method to design appropriate metrics, which are then applied to an example industrial project. Thus, they underline the common relevance and usefulness of measurement in software engineering and industry.

Flohr provides in [Flo08] a theoretical basis for quality gates as well as the design and definition of appropriate criteria. Quality gates are defined as particular milestones or decision points within a development process or project, which support the fulfillment of quality requirements, and which also map out a strategy with concrete quality objectives. He depicts when and how these criteria are identified from which actual metrics and how they can be improved over time.

## 3  Meta-Metrics for Software Quality Assurance

The term "meta-metric" for the development of vehicle functions is used to describe a methodical tool for providing necessary information to a vehicle project's stakeholders for example. Its application is required to steer and optimize the allocation of resources like developers, test engineers, or hardware-in-the-loop usage time and hence, to provide information about the quality of a certain development artifact and to maintain and improve its current level of quality. Therefore, various sources of information are continuously analyzed by mining data from the past over time for a single development artifact.

We define "meta-metrics" for simulations in software engineering in the automotive domain as follows: *The continuous determination of quantitative figures, which are defined over a set of results of simulation and test runs carried out for specific aspects, to steer and optimize the development process for an increased quality of the resulting product.*

In the following, several meta-metrics are defined to determine the quality of development artifacts over time. Hereby, $N$ describes the total number of individual and uniquely identifiable versions of a single development artifact; during a real development, $N$ might describe the total number of revisions from a centralized repository. In Eq. 1, first definitions are provided. Hereby, an artifact refers to either on a concrete software unit on a lower level or on an implementation model of an integrated vehicle function. The function $res(r,i)$ refers to an evaluation of a given test case, which might be a unit test or a complex traffic simulation model. $res(r,i)$ will be false iff a failure can only be assigned to the implementation side, else it will be true.

$$
\begin{aligned}
R_{succeeded}(r, N) &= \sum_{i=n_0}^{N} res(r, i) \\
\text{where } res(r, i) &= \begin{cases} 1 & \text{iff artifact } r \text{ was tested successfully at revision } i, \\ 0 & \text{else.} \end{cases}
\end{aligned}
\tag{1}
$$

$$
R_{failed}(r, N) = N - R_{succeeded}(r, N)
$$

Based on these initial definitions, the following first meta-metrics can be derived as described in Eq. 2, which describe the ratio of succeeding and failing development artifacts for the considered development period. By these meta-metrics, "heatmaps" may be generated, which visualize anomalies among the different development artifacts. Additionally, $Q_1$ compares the successfully carried out simulation runs for the development artifact $r$ during two development periods; if $Q_1$ is non-negative, a quality indicator can be derived to show that the quality of the considered development artifact has not decreased.

$$
\begin{aligned}
R^{+}(r, N) &= \frac{R_{succeeded}(r, N)}{N}, \\
R^{-}(r, N) &= 1 - R^{+}(r, N),
\end{aligned}
$$

$$Q_1(r, N_1, N_2) \;=\; R^+(r, N_2) - R^+(r, N_1) \text{ where } N_1 \leq N_2. \qquad (2)$$

To get an indicator about the long-term stability of a development artifact $r$, the following metrics are defined. First, the last revision is determined where the artifact $r$ failed during a test-run. Based on this revision number, its *negative* age is calculated describing how many revisions passed since the last failing one; thus, the larger this figure the better. The comparison of the ages for two different development periods as defined by $Q_2$ is an indicator whether the quality of $r$ has dropped during these two time points when $Q_2$ is negative.

$$\begin{aligned} \mathit{failed}(r, N) &= n \text{ with } n \in [0; N] \text{ where } n \text{ is the last failing revision for } r. \\ \mathrm{age}^-(r, N) &= N - \mathit{failed}(r, N) \end{aligned}$$

$$Q_2(r, N_1, N_2) \;=\; \mathrm{age}^-(r, N_2) - \mathrm{age}^-(r, N_1) \text{ where } N_1 < N_2. \qquad (3)$$

A further indicator for the average quality is the *mean time between test failures* (MTBTF). Obviously, the smaller the MTBTF value the worse is the quality and especially the reliability of the considered artifact $r$. In the following, $Q_3$ is defined to determine the MTBTF as shown in Eq. 4. The equation $R_{\mathit{failures}}(r, N)$ is used to determine the number of uniquely failing revision, i.e. consecutive failed revisions are considered as one failing revision unless the first non-failing revision is detected.

$$\begin{aligned} R_{\mathit{failures}}(r, N) &= \sum_{i=n_0}^{N-1} \mathit{fail}(r, i) \\ \text{where } \mathit{fail}(r, i) &= \begin{cases} 1 & \mathit{res}(r, i) \neq \mathit{res}(r, i+1) \wedge \mathit{res}(r, i) = 0, \\ 0 & \text{else.} \end{cases} \end{aligned}$$

$$Q_3(r, N) \;=\; \frac{R_{\mathit{succeeded}}(r, N)}{R_{\mathit{failures}}(r, N)}. \qquad (4)$$

In contrast to the aforementioned indicators, the following ones require access to the internal structure about an artifact's functionality to estimate its quality. The next indicator for an increased risk of potential failures is the inspection of the implementation model's complexity over time. A very naïve complexity function is the usage of the number of source code lines (SLOC); however, ineffectively written implementations may be considered as risky. Instead, there are better ways of describing the complexity and internal quality of the considered artifact $r$:

1. $sloc(r, i)$ This indicator describes the source lines of code artifact $r$ at revision $i$; this figure is required to relate the following indicators to it.

2. *MW(r, i)* This indicator describes the number of MISRA warnings at compile time for artifact $r$ at revision $i$ [MIS04].

3. *McC(r, i)* This indicator describes McCabe's complexity for artifact $r$ at revision $i$ [McC76].

4. *uncovered(r, i)* This indicator represents the number of uncovered statements during the execution of the simulation run.

All aforementioned indicators about an artifact's complexity or its internal quality can be embedded in the following meta-metric $Q_4(r, N_1, N_2)$ as shown in Eq. 5. Here, the meta-metric describes whether the average artifact's quality (i.e. its complexity or the number of critical MISRA-C warnings at compile time) has increased if it is a non-negative number.

$$Q_4(r, N_1, N_2) = \frac{\sum_{i=n_0}^{N_1} \frac{f(r,i)}{sloc(r,i)}}{N_1} - \frac{\sum_{i=n_0}^{N_2} \frac{f(r,i)}{sloc(r,i)}}{N_2} \quad (5)$$

where $0 < N_1 \leq N_2 \wedge f(r, i)$ is one of the aforementioned functions.

The next meta-metrics require the measurement of execution time and acting time points during the simulations. In Eq. 6, the indicator $Q_5(r, N_1, N_2)$ is defined which relates the average execution time in the simulation for artifact $r$ for development period $N_1$ to the period $N_2$. Hereby, it is assumed that the execution time in the simulation is only dependent from modeled situation and from the underlying simulation engine. Thus, a non-negative $Q_5(r, N_1, N_2)$ might indicate an increased performance of the algorithm for instance.

$$duration(r,i) = \begin{cases} \text{required execution time in the simulation for artifact} \\ \quad r \text{ at revision } i \text{ iff } res(r,i) = 1 \\ 0 \text{ else.} \end{cases}$$

$$Q_5(r, N_1, N_2) = \frac{\sum_{i=n_0}^{N_1} duration(r,i)}{R_{succeeded}(r, N_1)} - \frac{\sum_{i=n_0}^{N_2} duration(r,i)}{R_{succeeded}(r, N_2)} \quad (6)$$

where $0 < N_1 \leq N_2$.

In Eq. 7, the indicator $Q_6(r, N, s)$ determines the population standard deviation of the acting time point for the artifact $r$ during the development period $N$. Thus, it can be quantitatively determined how reliably an algorithm is acting within a given situation $s$ in the simulation over time: A negative $Q_6(r, N_1, N_2, s)$ reflects that the variance in an algorithm's acting time point has increased and thus, the overall quality of the algorithm has decreased for artifact $r$ wrt. the specific situation $s$.

$$acting(r,i,s) = \begin{cases} \text{acting time point for the situation } s \text{ for artifact} \\ \quad r \text{ at revision } i \text{ iff } res(r,i) = 1 \\ 0 \text{ else.} \end{cases}$$

$$\overline{acting}(r, N, s) = \frac{\sum_{i=n_0}^{N} acting(r, i, s)}{R_{succeeded}(r, N)}$$

$$v(r, N, s) = \sqrt{\frac{\sum_{i=n_0}^{N} (acting(r, i, s) - \overline{acting}(r, N, s))^2}{R_{succeeded}(r, N)}}$$

$$Q_6(r, N_1, N_2, s) = v(r, N_1, s) - v(r, N_2, s) \tag{7}$$

## 4 Application on Testing Procedures

The development of an automotive safety function is shaped by a lot of individual activities consisting of quite a number of different, concurrently fulfilled tasks. Thus, several process models exist to structure those activities into single steps to define important milestones and quality gates, and to offer guidance, e.g. methods, specific roles and responsibilities and the like. Established models are the V-model or its more elaborated W-model variant, but also evolutionary models such as rapid application/product development (RAD/RPD), extreme programing (XP) or dynamic systems development method (DSDM) as mentioned in [SRWL11].

The characteristics of automotive systems therein are that any development of a system stretches not only on a software product but also the underlying hardware. Those device controllers are often designed for that particular software or vice versa, which leads to varying test activities as well, e.g. communication behavior, arithmetic speed, and the like. Subjective and objective evaluation methods provide the basis for quality measurement during the test process at the various abstraction layers. Therefore, an ideal-typical development and testing process may look like as in Fig. 1.

This development and test model may be complemented by an iterative or incremental view: Every quality gate defines its own requirements and test cases to realize a specific range of that function. The principle of continuous integration, which means the periodic integration of every single line of code and its corresponding test case at a centralized repository, as well as the test execution refines this model additionally and leads to a nearly freely selectable granularity. Every single revision represents an evaluable function to a specific moment in time of the project progression. In that context, meta-metrics can be identified and applied.

The aforementioned meta-metrics enable to derive statements about the relative quality change of that function or system or an indication where the quality might have changed during a specific part of the development process. The outlined approach aims for supporting project managers and engineers by a method that is able to reveal possible quality anomalies in the function's realization or implementation without requiring to know explicitly technical details of the source code. This circumstance can be found often in a distributed development setting where systems are realized by an OEM and its 1-tier suppliers.

To get familiar with using these meta-metrics, they are firstly explained exemplarily to show how they work and how they are measured. In a second step, the meta-metrics are being projected on the field of the development of automotive safety functions to show their applicability and usability. Finally, it is shown at which certain point of the process these meta-metrics can support the development of ISS.

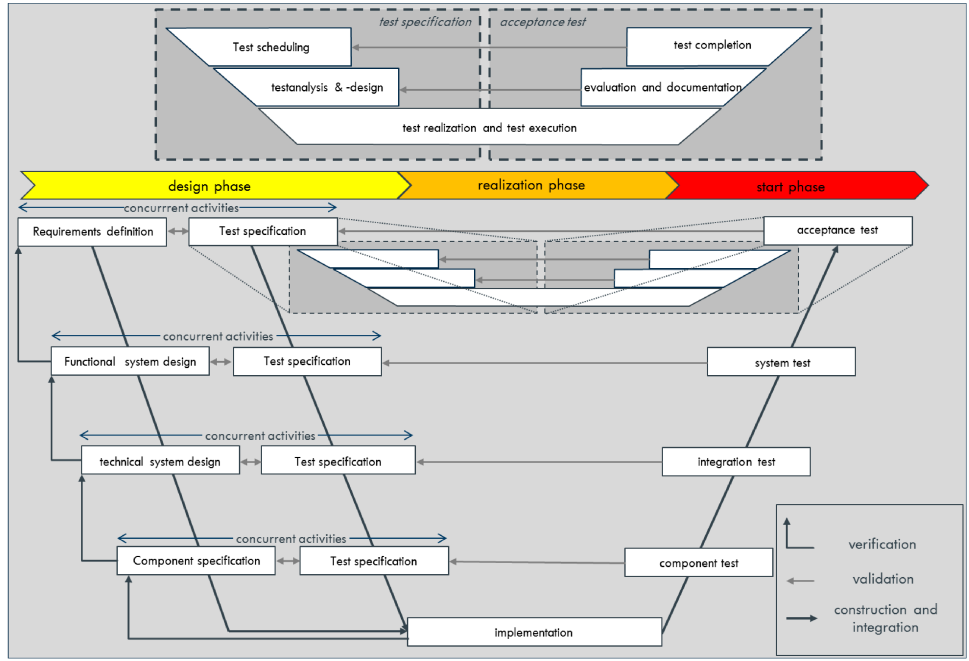

Figure 1: V-model: An example for a development process referring to [SRWL11].

**Exemplary application**  To illustrate the functionality of the aforementioned meta-metrics, they are explained by a revision history of a simulation framework, which was used during the development of an autonomous ground vehicle (AGV) [Ber10]. The software revision history documents 1,867 single revisions for all components. The further interest focuses on the artifact "AGV communication and control software" (CCS), which has 892 separate revisions representing changes to the internal code structure with an accompanying test-runs. The application of the aforementioned meta-metrics resulted in the following figures:

$R_{succeeded}(\texttt{CCS}, 892) = 192$

$R_{failed}(\texttt{CCS}, 892) = 700$

In regard the relative ratios are:

$R^{+}(\texttt{CCS}, 892) = 0.2152$

$R^{-}(\texttt{CCS}, 892) = 0.7848$

With respect to two revisions in the progression of the development process, a positive quality change is indicated as follows. The revision no. $N_1$ und $N_2$ constitute particular defined quality gates, so each $R^+$ describes the relative frequency of succeeded results during each development cycle from the beginning to the regarding quality gate. The last failing revision no. and its related age are:

$\mathit{failed}(\mathtt{CCS}, 892) = 743$

$\mathrm{age}^-(r, N) = 149$.

It follows with

$Q_2(\mathtt{CCS}, 768, 892) = \mathrm{age}^-(\mathtt{CCS}, 892) - \mathrm{age}^-(\mathtt{CCS}, 768) = 149 - 25 = 124$

as an indicator, which reveals that there is no negative change of quality during the development of that artifact. With an acceptance level of 100% success rate of all tests the mean time between test failures represented by $Q_3$ resulted in

$Q_3(\mathtt{CCS}, 892) = \frac{R_{\mathit{succeeded}}(\mathtt{CCS}, 892)}{R_{\mathit{failures}}(\mathtt{CCS}, 892)} = \frac{192}{3} = 64,$

so after a period of 64 successful revisions, it is likely of having a failing revision. Usually, the MTBTF should increase here while the maturity of the software is rising, too.

The other meta-metrics can be calculated as mentioned in Sec. 3 in an analog manner and are left out for the sake of simplicity.

**Which benefits may arise from these meta-metrics for the development of ISS?** Because of the different abstraction levels of the V-model and the refined requirements between the separate layers, logical and technical details of a function are broken down continuously. Thus, the respective engineers and developers are integrated in technical aspects differently. On the one hand, this derives various views for a developer on the total function, which is actually intended to apply the more-eye-principle and to reveal a failure more directly and earlier in the development phase. On the other hand, however, it means that a part of the participants involved into the project has a varying knowledge of the technical aspects depending on the level of abstraction. In this particular case, the presented meta-metrics can provide assistance with estimating the quality of software that is tested both in a simulation environment and in an extension of real test drives. From that point, further steps can be taken for improvement.

In addition to already established methods, the engineer is enabled to estimate the quality of the corresponding software by these meta-metrics as indicators without the need of having a detailed knowledge about the underlying implementation. Especially in a simulation environment, in which the source code can be integrated as a test object for software- or hardware-in-the-loop tests, meta-metrics $Q_1$ to $Q_6$ can unveil possible anomalies within the software module even on a higher level of abstraction.

Furthermore, they can also be applied in association with suppliers who develop a function by a contractual dependency with an OEM because hereby, an evaluation is possible while retaining the supplier's intellectual property. Therefore, these metrics are also suitable for blackbox testing, as far as both sides are able to achieve an agreement of providing the necessary data for the calculation. Is there an arrangement to provide iteratively a specific range of functions, the presented metrics can be applied to those milestones or quality gates. At least, those figures would be only available as the lowest level of granularity under the aforementioned circumstances. If the function is an truly in-house development with no other suppliers wrt. to the software, quality statements can be generated on a daily base. Meta-metrics also provide support in course of functional safety and accordance with ISO-26262 where OEMs and suppliers have to accomplish their documentation obligation while developing ISS for new vehicles.

Future generations of simulation environments will have a significant impact on the development of ISS and will offer new opportunities to reduce the risk of against failure and for designing a function. Here, a similar evolution is conceivable such as finite-element-method (FEM) and multi-body-simulation (MBS) carried out in vehicle body construction regarding safety aspects. Therefore particular tests are firstly executed in a virtual environment before these tests are verified and validated in reality to reduce the risk of failure as well. The riskiness of such situations for people and equipment in context of ISS like potential and concrete crash scenarios will increase the relevance of those virtual environments. The benefits of repeatability as well as persistent availability will lead to a further increase of attractiveness in functional development of ISS. The need of automation of such environments will offer new testing methods to engineers and developers like continuous integration did in software-engineering.

Hence, meta-metrics will have great potential to handle the large amount of data more easily, which are generated by simulation test runs in the context of functional development especially when it comes to estimating the quality changes within time-critical safety functions. For instance, each delayed acting time point for an emergency braking may cause hazardous consequences for all involved people. $Q_5$ und $Q_6$ are now able to provide specific information as an indicator of quality changes regarding the timing behavior of an algorithm at certain quality gates, when the set of simulation scenarios and parameters remains the same, though. If a delay in acting time points or even in the duration of test execution is being detected over the development time, the source code may have changed in a negative manner and further investigation should be made.

The following Fig. 2 reveals, at which particular points in the common V-model the aforementioned meta-metrics may be used.

## 5 Conclusion and Future Work

To increase the traffic safety for both the driver and other participants, ISS become continuously more important; so the necessary software to realize such systems increases in complexity and cross-linkage as well. Thus, ongoing enhancement in testing methods, procedures, and supporting tools is strongly recommended to fulfill safety requirements and the customer's requests. This article presents a contribution aiming to support engineers and developers during the development by appropriate metrics measuring software and functions.

Especially meta-metrics may help to make a detailed evaluation of existing metrics at functions and software modules possible. Thus, a separate approach has been adopted to characterize firstly meta-metrics in an automotive context and to derive an appropriate definition for this contribution afterwards. After that, concrete meta-metrics have been presented, which operate as an indicator to estimate the quality of an ISS over time.

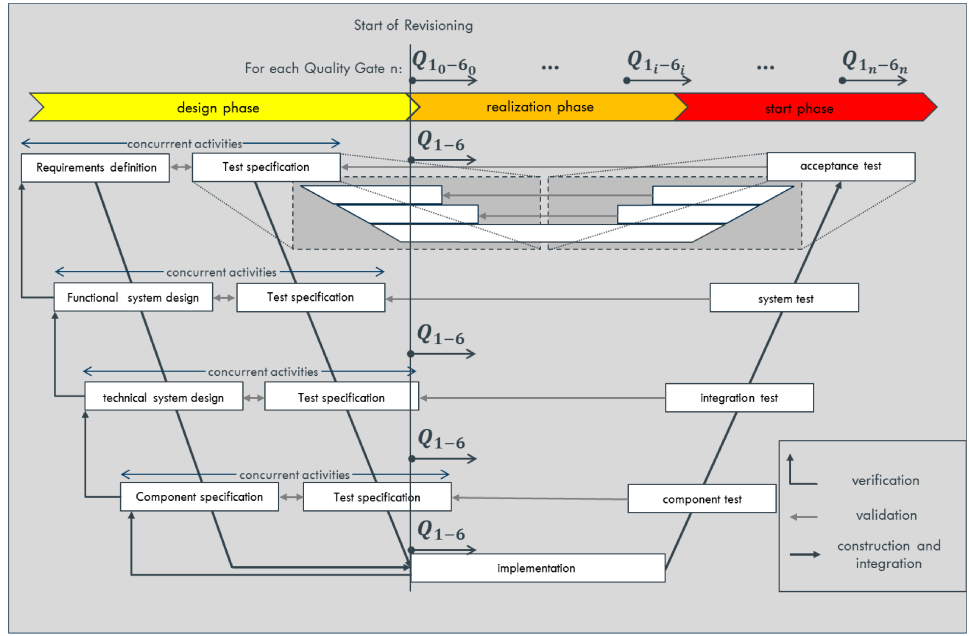

Figure 2: V-Model: Potential application areas during the V-model.

Finally, several meta-metrics have been expounded for a more intuitive understanding and how they may be helpful within the development process of ISS and its testing procedures. It could be shown that meta-metrics may indicate quality changes at certain milestones or agreed quality gates during a software project. Because of their independence of the technical context, they can be applied on different abstraction levels within the common V-model.

Future work will concentrate on further proving the presented metrics in certain development projects and focusing especially on those with using simulative approaches as testing procedures. More research in the field of meta-metrics will be done as well to derive new meta-metrics in the automotive context and to develop best practices as well.

# References

[Aly06] Vadym Alyokhin. Management von Softwaresystemen - Systembewertung: Metriken und Prozess. http://www4.in.tum.de/lehre/seminare/hs/WS0506/mvs/files/Ausarbeitung_Alyokhin.pdf, 2006.

[BBeA03] Aline Lucia Baroni and Fernando Brito e Abreu. A Formal Library for Aiding Metrics

| | Extraction. In *ECOOP Workshop on Object-Oriented Re-Engineering, Darmstadt, Germany*, 2003. |
|---|---|
| [Ber10] | Christian Berger. *Automating Acceptance Tests for Sensor- and Actuator-based Systems on the Example of Autonomous Vehicles*. Shaker Verlag, Aachener Informatik-Berichte, Software Engineering Band 6, Aachen, Germany, 2010. |
| [Ber12] | Christian Berger. From Autonomous Vehicles to Safer Cars: Selected Challenges for the Software Engineering. In *Proceedings of the Conference Automotive - Safety & Security*, pages 1–10, Magdeburg, Germany, September 2012. |
| [BR12a] | Christian Berger and Bernhard Rumpe. Autonomous Driving - 5 Years after the Urban Challenge: The Anticipatory Vehicle as a Cyber-Physical System. In Ursula Goltz, Marcus Magnor, Hans-Jürgen Appelrath, Herbert K. Matthies, Wolf-Tilo Balke, and Lars Wolf, editors, *Proceedings of the INFORMATIK 2012*, Braunschweig, Germany, September 2012. |
| [BR12b] | Christian Berger and Bernhard Rumpe. Engineering Autonomous Driving Software. In Christopher Rouff and Mike Hinchey, editors, *Experience from the DARPA Urban Challenge*, pages 243–271. Springer-Verlag, London, UK, 2012. |
| [BRR10a] | Christian Berger, Holger Rendel, and Bernhard Rumpe. Measuring the Ability to Form a Product Line from Existing Products. In *Proceedings of the Fourth International Workshop on Variability Modelling of Software-intensive Systems (VaMoS)*. University of Duisburg-Essen, 2010. |
| [BRR$^+$10b] | Christian Berger, Holger Rendel, Bernhard Rumpe, Carsten Busse, Thorsten Jablonski, and Fabian Wolf. Product Line Metrics for Legacy Software in Practice. In *Proceedings of the 14th International Software Product Line Conference (SPLC 2010) Volume 2*. Lancester University, 2010. |
| [EFR07] | Irene Eusgeld, Felix C. Freiling, and Ralf Reussner. *Dependability Metrics: Advanced Lectures*. Springer, 2007. |
| [Flo08] | Thomas Flohr. Defining Suitable Criteria for Quality Gates. In *Proceedings of the International Conferences on Software Process and Product Measurement*, IWSM/Metrikon/Mensura '08, pages 245–256. Springer-Verlag, 2008. |
| [FP98] | Norman E. Fenton and Shari Lawrence Pfleeger. *Software Metrics: A Rigorous and Practical Approach*. PWS Publishing Co., 2nd edition, 1998. |
| [Glo05] | Wolfgang Globke. Software-Metriken. http://www.math.kit.edu/iag2/ globke/seite/seminar/media/metriken.pdf, Juni 2005. |
| [LDA97] | Franz Lehner, Reiner Dumke, and Alain Abran. *Software-Metrics: Research and Practice in Software Measurement*. Dt. Univ.Verlag, 1997. |
| [Lig09] | Peter Liggesmeyer. *Software-Qualität - Testen, Analysieren, Verifizieren von Software*. Spektrum-Verlag, 2009. |
| [McC76] | Thomas J. McCabe. A Complexity Measure. *IEEE Transactions on Software Engineering*, SE-2(4):308–320, December 1976. |
| [MIS04] | MISRA. *MISRA-C:2004 - Guidelines for the use of the C language in critical systems*. Motor Industry Research Association, 2004. |
| [NCA12] | Euro NCAP. Euro NCAP to drive availability of Autonomous Emergency Braking systems for safer Cars in Europe. Press Release on website, June 2012. |

[OMG12]   OMG. Unified Modeling Language Specification, Version 2.4.1. Technical report, Object Management Group, 2012.

[SRWL11]  Andreas Spillner, Thomas Rossner, Mario Winter, and Tilo Linz. *Praxiswissen Softwaretest - Testmanagement*. dpunkt.verlag, 2011.

[SSB10]   Harry M. Sneed, Richard Seidl, and Manfred Baumgartner. *Software in Zahlen - Die Vermessung von Applikationen*. Hanser-Verlag, 2010.

[SX09]    A. Stefani and M. Xenos. Meta-Metric Evaluation of E-Commerce-related Metrics. *Electronic Notes in Theoretical Computer Science*, 233:59–72, mar 2009.

[WN10]    Edzard Weber and Andre Nimmich. Meta-Kennzahlen fuer die Bewertung von Dienstleistungen. In *Diskussionsbeitraege des 2. Workshops Dienstleistungsmodellierung (DLM 2010)*, pages 65–88, 2010.

[Woo99]   Terry L. Woodings. Meta-Metrics for the Accuracy of Software Project Estimation. http://citeseerx.ist.psu.edu/viewdoc/download?doi=10.1.1.118.2390, 1999.

[XSZC00]  M. Xenos, D. Stavrinoudis, K. Zikouli, and D. Christodoulakis. Object-Oriented Metrics - A Survey. In *Proceedings of the FESMA Conference (FESMA'2000)*, 2000.